\documentclass[12pt]{article}          
\usepackage[draft]{hyperref} 
\usepackage{graphicx}

\newcommand\Softitle[1]{\Large \bf \noindent \begin{center} #1
\end{center}\rm \normalsize \vskip.125in }%

\newcommand\Sofauthor[1]{\vskip.1in\noindent%
   \large  \begin{center} \textsf{#1} \end{center}\rm \vskip-.2in}

\let\title\Softitle
\let\author\Sofauthor

\let\address\Sofaddress
\let\email\Sofemail

\begin{document}

\title{Characterisation of pulsed Carbon fiber illuminators
       for FIR instrument calibration}

\author{S. Henrot-Versill\'e, R. Cizeron, F. Couchot}

\address{LAL B\^{a}t 200}

\email{versille@lal.in2p3.fr, cizeron@lal.in2p3.fr, couchot@lal.in2p3.fr}

\begin{abstract}
We manufactured pulsed illuminators emitting in the far infrared for the Planck-HFI bolometric
instrument ground calibrations. Specific measurements have been conducted on these light sources,
based on Carbon fibers, to understand and predict their properties.
We present a modelisation of the temperature dependence of the thermal
conductivity and the calorific capacitance of the fibers.
A comparison between simulations and bolometer data is given, that shows the coherence of
our model.
Their small time constants, their stability and their emission spectrum
pointing in the submm range make these illuminators a very usefull tool for
calibrating FIR instruments.
\end{abstract}



\section{Introduction}
We have been studying Carbon fibers as illuminators for FIR instrument calibration between 2000 and
2005 with the specific goal of measuring the sum of the electrical and optical crosstalk
\footnote
{The optical crosstalk is caused by potential leakages between the different
 cryogenic stages of the optics facing the bolometers.}
 of the Planck-HFI instrument\cite{3}\cite{HFI}. Our requirements were threefold:
\begin{itemize}
\item{}  to get a signal ranging up to a few pW detected on the Planck-HFI bolometers for the concerned
frequencies (ranging from 100 to 857 GHz) including transmission efficiency of the cold optics horns
and integration over the $30\%$ wide frequency bands.
\item{} to get a relatively small time constant (smaller than 10ms) and a reproducible
signal in a pulsed regime, allowing to stack the measurements in order to increase the
signal over noise ratio.
\item{} to be able to "lighten" the fibers without introducing any parasitic 
signal due to EMI-EMC from the electrical pulse used to drive them.
\end{itemize}
We did fulfill these requirements. 
Even more, on the time constant issue, we did get better than expected results and the fibers
were one of the sources used to characterise the time response of the instrument.

Since the carbon fibers are composite materials, no measurement of the particular properties of the ones
we used have been made. We therefore focus in this paper on their characterisation.
In part 2 we give the heat equation and the corresponding approximations made in this paper to
analyse the fiber behaviour. Part 3 describes the experimental setup and in part 4 we present the 
measurements of the relation between resistance and temperature and the analysis that leads to the
measurement of the thermal conductivity and the calorific capacitance. Part 5 describes the fiber modelisation
and a comparison of simulation results with data.
\section{General framework}
\label{frame}
\subsection{Equations and assumptions}

\begin{figure}[htbp]
\centering
\includegraphics[width=6.0cm]{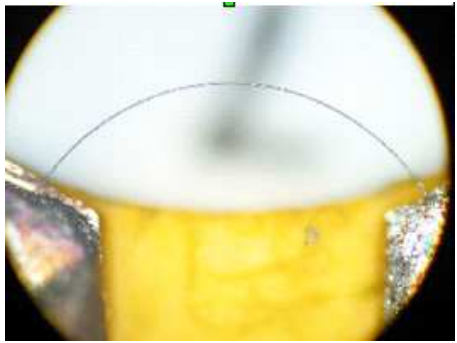}
\caption{\label{photo} {Photography of one of the carbon fiber used for the Planck-HFI calibration:
the diameter is 6$\mu$m and the length $\simeq 1$ mm.}}
\end{figure}

The modelisation of the carbon fiber behaviour relies on the one dimensionnal heat equation
(along the fiber):
\begin{eqnarray}
\label{maineq}
{\partial \over \partial x} ( \kappa S{ \partial  T \over \partial x} ) + {\mu I^2\over S} = \rho C_p S
{\partial T \over \partial t}
\end{eqnarray}

\noindent where:
\begin{itemize}
\item{}$\kappa$ is the thermal conductivity, $C_p$ the calorific capacitance, $\mu$ the resistivity.
We assume here that all these parameters do depend on temperature $T$.
\item{} $\rho$ the mass per unit of volum, and S the surface of a slice \cite{4}. The resistance $R$
is obtained through $\mu(T)$ integration along the length $L$ of the fiber.
For the fibers we used, $L\simeq 1 mm$ and the diameter of a slice is 6 microns: a photography is
shown on figure \ref{photo}. This picture also shows the way it is thermalised at each end
on a kapton circuit through Ag lacquer contacts.
\item{} The fiber is powered by a current (I) generator.
\item{} The radiated power represents less than 1\% of thermal losses, even in the worst case
\footnote{The fiber radiates about $10 \mu W$ when heated at
 about 200K by $1 V$ pulses, dissipating about $1 mW$ Joule power.},
 so it is neglected with respect to conduction.
\end{itemize}
\subsection{Basic properties}
The fiber is thermalised at each end at a temperature $T_0$ (cf. figure \ref{photo}).
In the steady state regime, and in the simplest case where no parameter exhibits
 any temperature dependence, Eq. (\ref{maineq}) gives:
\begin{eqnarray}
\label{simpeq}
\kappa{ \partial^2  T \over \partial x^2}  + {RI^2\over LS}  = 0
\end{eqnarray}
leading to (the fiber center being the $x$ axis origin): 
\begin{eqnarray}
\label{tx}
T(x) = T_0 + { RI^2 \over 2 L S \kappa  } ( { L^2 \over 4 } - x^2 ) \ .
\end{eqnarray}

\noindent {and a mean temperature increase:}
\begin{eqnarray}
\label{deltmes}
\Delta T_{mes} &=& {1\over L}\int_{-L/2}^{L/2} (T(x)-T_0) dx
= {R I^2 L\over 12 \kappa S} \ .
\end{eqnarray}

When one stops heating the fiber, the system relaxes to the $T_0$ temperature,
and the transcient regime can be described by the generic Fourier expansion
solution:
\begin{eqnarray}
T(x,t)&=&T_0+\sum_{k=0}^\infty \Delta T_0^k e^{(-t/\tau_k)}  \cos( (2k+1)\pi x /L) \ ,
\end{eqnarray}
where the time constants $\tau_k$ are given by:
\begin{eqnarray}
\label{tauk}
\tau_k&=& {\rho C_p L^2 \over \kappa (2k+1)^2 \pi^2} \ ,
\end{eqnarray}
the leading term being by far the $k=0$ component, which corresponds to the longest time constant:
\begin{eqnarray}
\label{tau1}
\tau_0 &=& {\rho C_p L^2 \over \kappa  \pi^2} \ .
\end{eqnarray}
The fiber time constant is then proportional to the square of its length. Therefore, smaller time constants may be obtained with shorter fibers, at the price of a flux loss. $L$ should stay sizeable with the longest needed wavelength, since the fiber acts like an antenna.

For small values of $I$, $\Delta T_{mes}$ is small, and these formula apply even if $\kappa$ varies with $T$.
We use this approximation in the following.

Resistivity dependence with T can be treated analytically in the simplest case where $\kappa(T)$
is constant, assuming $\mu(T)$ varies linearly as:
\begin{eqnarray}
\label{mualp}
\mu(T)=\mu_0\left[ 1-\alpha (T-T_0)\right] \ ,
\end{eqnarray}

\noindent then, Eq. (\ref{maineq}) exact solution for the mean temperature increase in steady state regime
reads\cite{4}:

\begin{eqnarray}
\Delta T_{mes} &=& {1\over \alpha} \left[1 - {1\over {L I\over 2 S} \sqrt{{\mu_0 \alpha\over
\kappa}}}
 \tanh\left({L I\over 2 S}\sqrt{{\mu_0 \alpha\over \kappa}}\right)\right] \ .
\end{eqnarray}

\section{Experimental setups}
\label{Exper}

The measurements described in the following sections have been performed
 within the Saturne cryostat\footnote{This cryostat at IAS-Orsay
has been used for the calibration of the Planck-HFI instrument.} providing us
with a thermalisation temperature ranging from 300K to 1.7K
(we took advantage of the cooling down of the setup).
Two kind of data are analysed: 
\begin{itemize}
\item{} resistance (or resistance variations) measurements at the edges of the fibers done using the dedicated
electronics described below for different thermalisation temperatures and for small values of the current
 applied on the fibers.
\item{} bolometers' data translated in terms of incident power for a
thermalisation temperature around 2K and a wide range of current values.
\end{itemize}

\subsection{Dedicated electronics description}
\label{descr_dedi}
\begin{figure}[htbp]
\centering
\includegraphics[width=7.0cm]{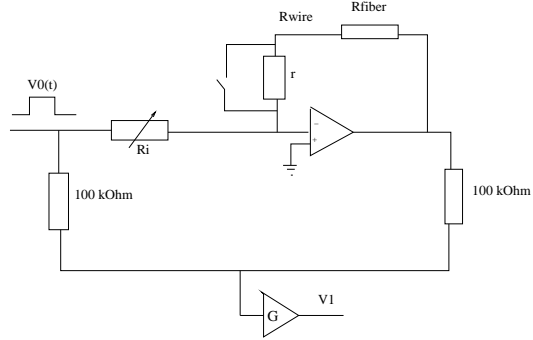} 
\caption{\label{Elecfibre} {Figure illustrating the electronic circuit used
to characterise the resistance of the carbon fibers.}}
\end{figure}

\begin{figure}[htbp]
\centering
\includegraphics[width=6.5cm]{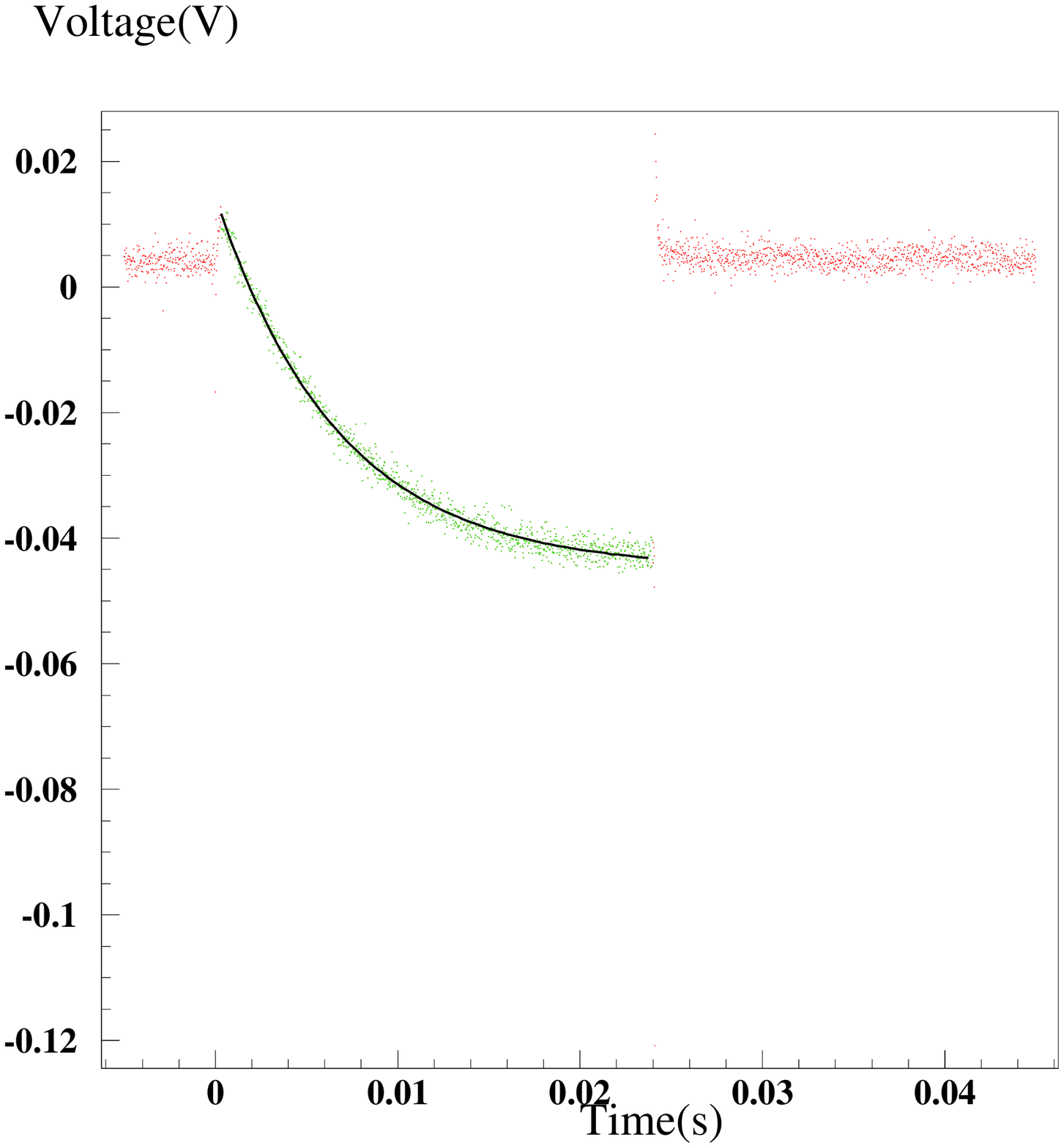}
\includegraphics[width=6.5cm]{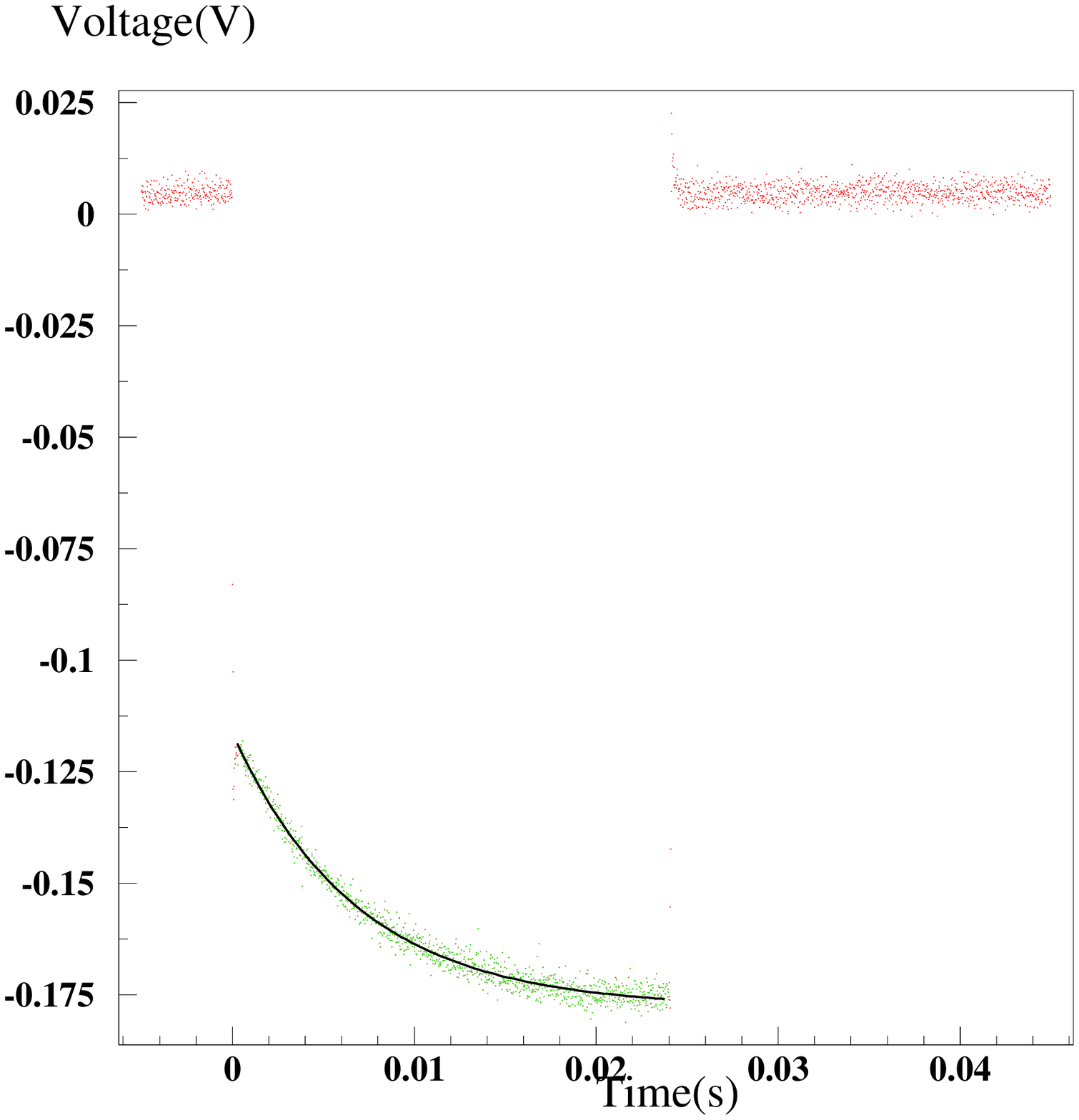}
\caption{\label{Elecfibre2} { Example
of $V_1(t)$ when a square signal $V_0$ of $0.2\ V$ is applied: on the left when 
$r (\simeq 2.6 \Omega) $ is switched on and on the right when it is off.
The device reaches a precision of the order of $.05 \Omega$ on the resistance variation
measurements. }}
\end{figure}

The electronics used for resistance and resistance variation measurements is described
in figure \ref{Elecfibre}.
The signal $V_0$ entering the circuit is a 
low impedance square signal which amplitude, frequency, and 
width can be changed. The output is simply given by: 
\begin{eqnarray}
V_1(t)  =  ( 1-{\tilde{R}_{tot}(t)\over R_i}) G V_0(t)
\end{eqnarray}
where $\tilde{R}_{tot}$ corresponds to  the resistance of the fiber in series with
 15 meters long wires.
Before any measurement we balance the system in order to get:
$R_i\simeq \tilde{R}_{tot}(t=0) $.
Taking $t=0$ at the beginning of the pulse, one gets:
\begin{eqnarray}
V_1(t)  = V_1(0)+ {G V_0(t)\over R_i} (R_{fiber}(t)-R_{fiber}(0)) 
\end{eqnarray}
The gain factor is measured by switching on and off the small additionnal resistor $r$.  

An example of the signal measured at the edges of the fibers is illustrated
on  figure \ref{Elecfibre2} (cf. section \ref{dedi} for more details).

\subsection{Saturne setup}
\begin{figure}[htbp]
\centering
\includegraphics[width=6.0cm]{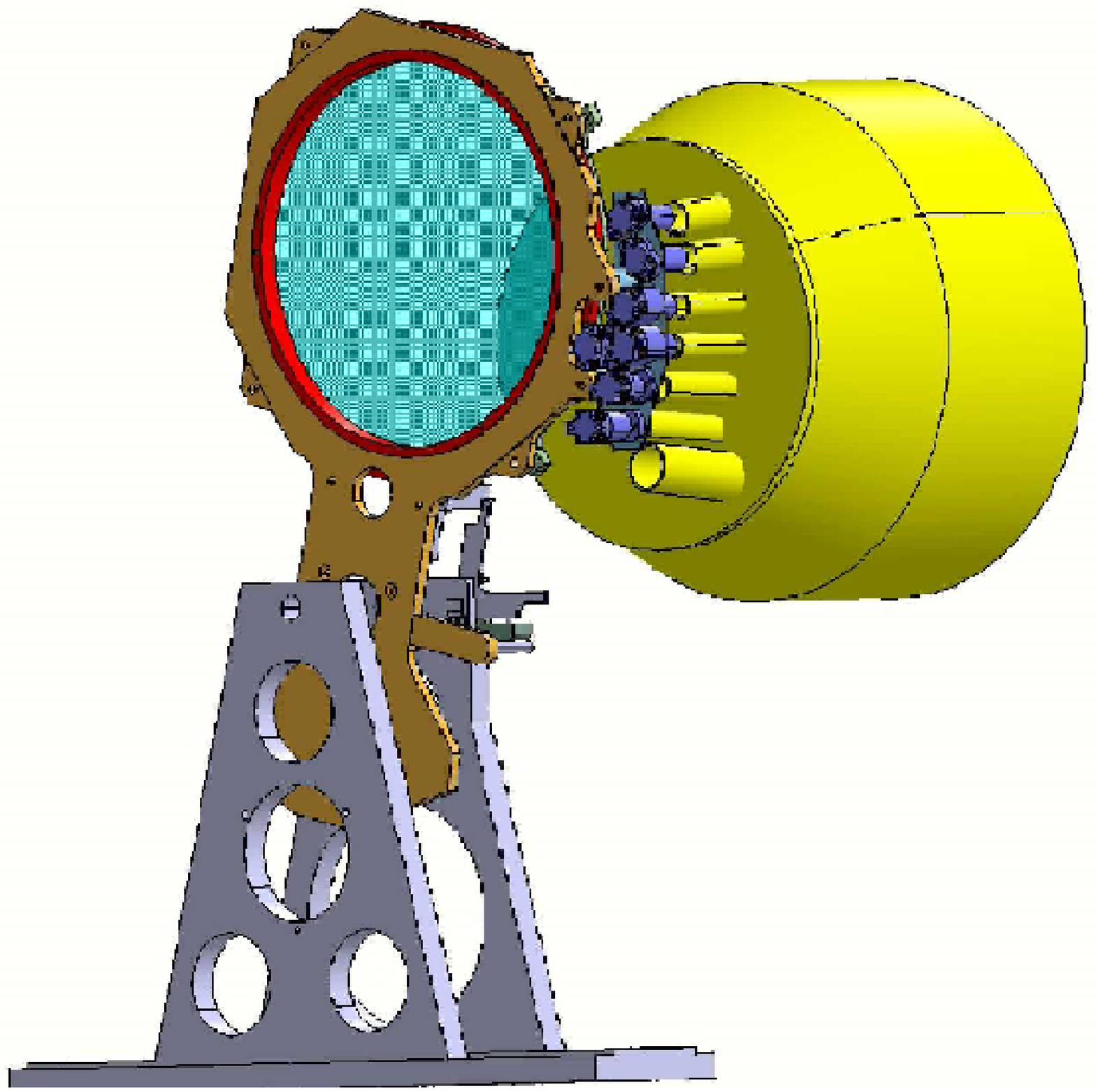} 
\includegraphics[width=4.0cm]{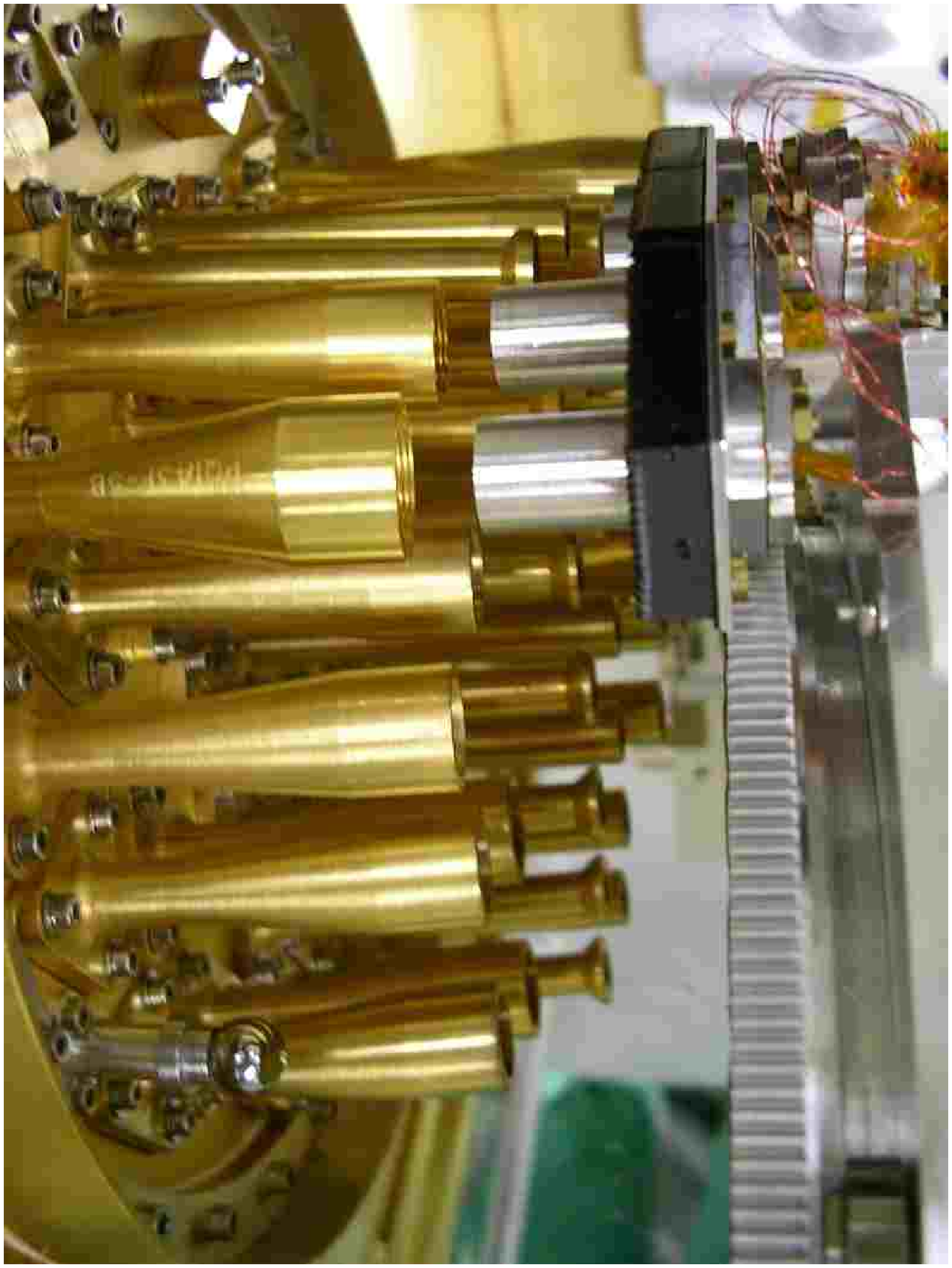} 
\caption{\label{setup} {Schematics and picture of the setup used for the
characterisation of the Planck-HFI instrument
(which cold optics is represented in yellow on the left and which is in gold
on the picture) with the carbon fibers illuminators (resp. in blue and in grey).}}
\end{figure}

The setup of the fibers within the Saturne cryostat did take into account the
constraints of all the measurements which were needed to be done to calibrate
the Planck-HFI instrument (and which are not discussed here\cite{Calib}).
The fibers were therefore installed on a side of a 3 positions wheel:
for one of them, the fibers were facing the entrance of part of the cold optics of
the instrument as shown on figure \ref{setup}. Two additionnal fibers were installed behind a 
hole in a mirror facing the focal plane and illuminating all the bolometers synchroneously, for time
constant measurements. Being further from the instrument, the mirror fibers illuminate HFI horns about
$300$ times less than the wheel fibers. 

\section{Characterisation}
\label{carac}
\subsection{Resistance and temperature}
\label{Rdet}
\begin{figure}[htpb]
\begin{center}
\includegraphics[width=7.cm]{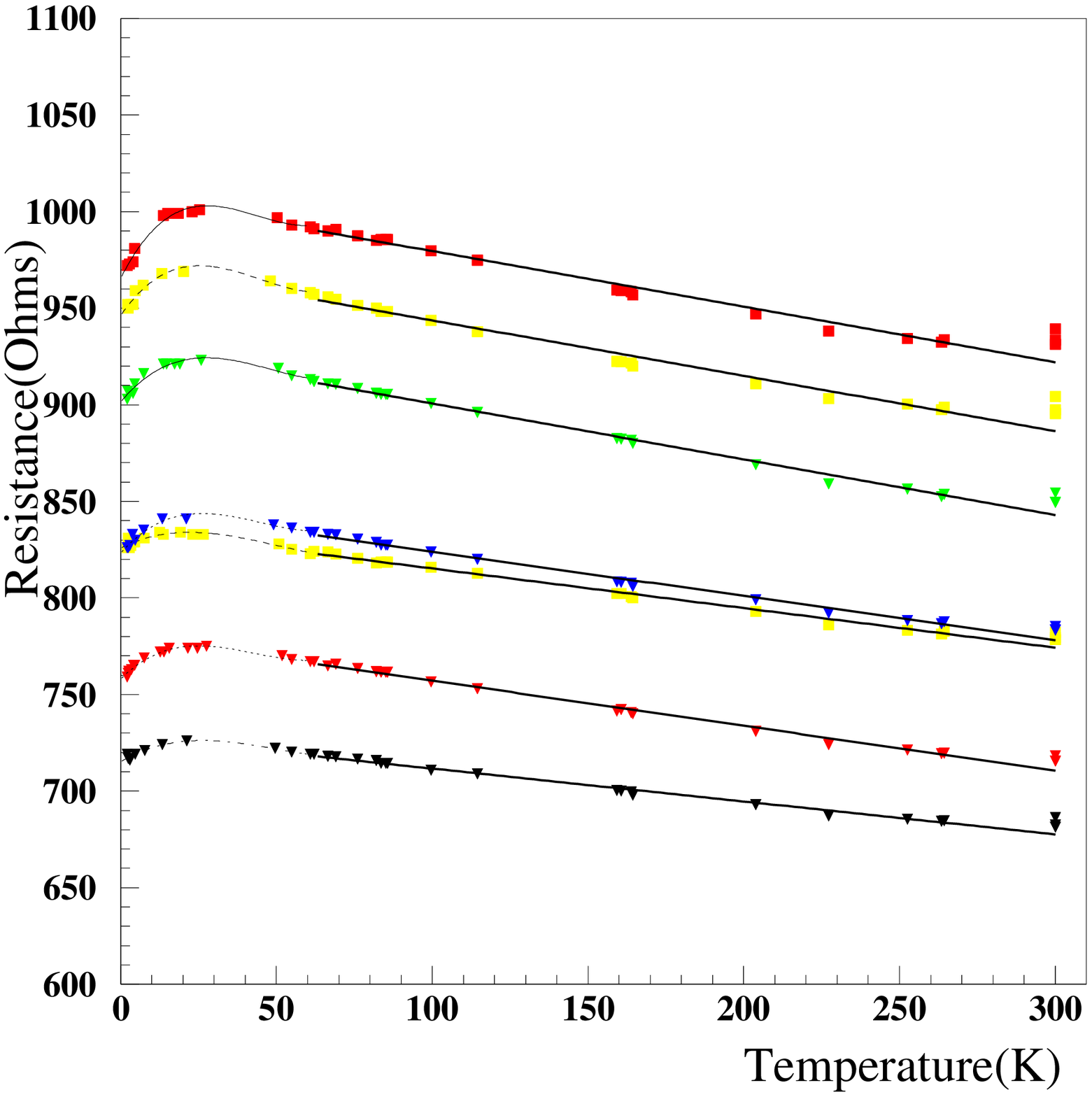}
\includegraphics[width=7.cm]{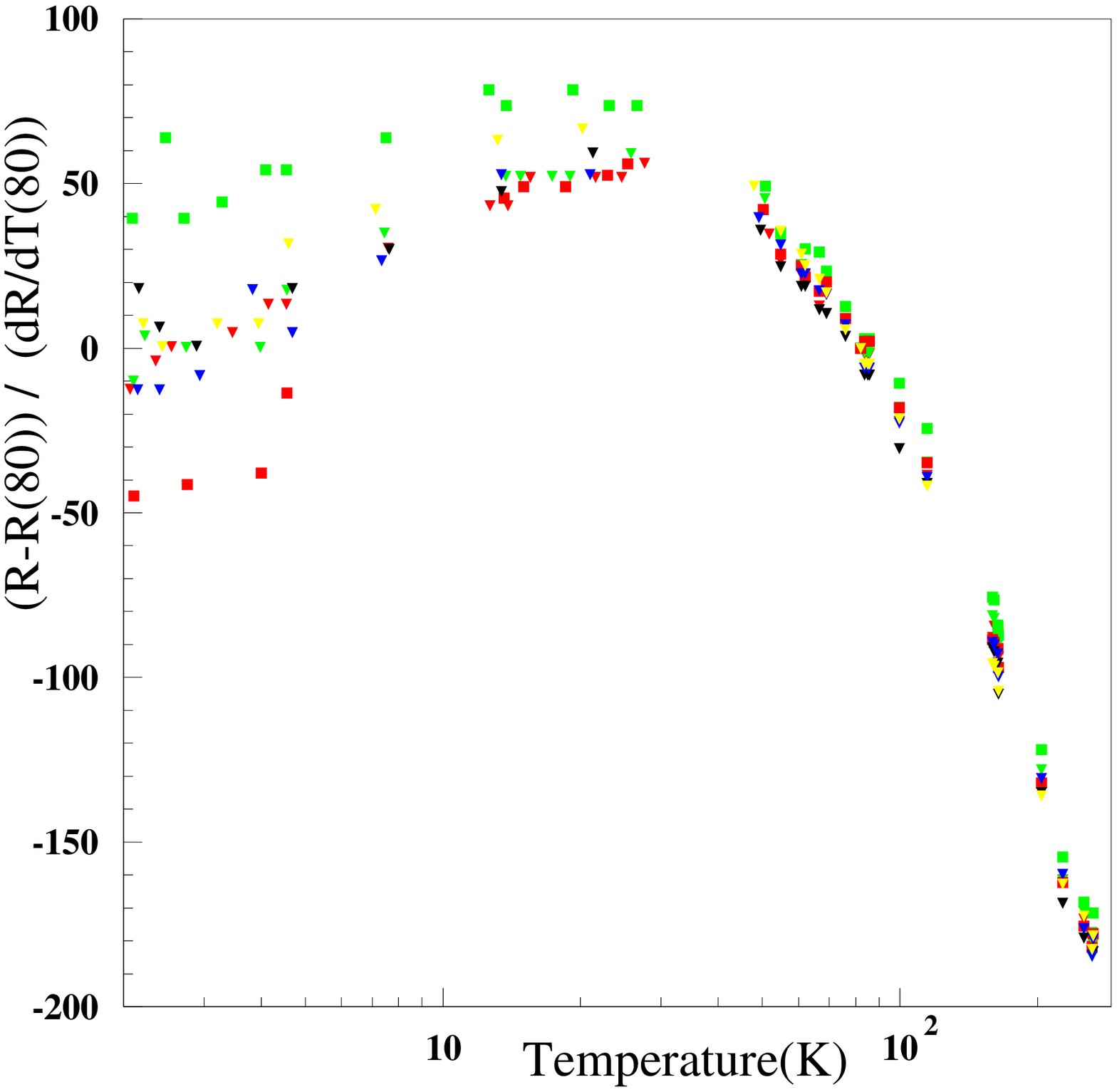}
\caption{\label{temptherm} On the left: resistance of the fibers as a function of 
the thermalisation temperature (a polynomial fit is superimposed). On
the right: comparison of the same data when properly normalized (cf. text).}
\end{center}
\end{figure}

At room temperature the measured resistivity of the fibers
is $1.8 \ 10^{-5} \Omega . m$ which gives for 1mm fibers
a resistance of 640 Ohms at $300 K$. This resistance does depend on the thermalisation
temperature as shown on figure \ref{temptherm} for seven fibers (on the left): 
the dispersion of these measurements includes the spread of the resistance of
the Ag-lacquer contacts  and the fact that the length of the fibers lies between
0.94 and 1.06 mm. The figure on the right shows, for the same seven fibers,
their resistance corrected for their value measured at 80K and divided
by the slope ${d R\over d T}$ at $T=80 K$ to correct for length effects. It 
illustrates the homogeneity of the results. The contact resistance
due to the Carbon-Ag lacquer interfaces is of the order of $100 \Omega$.

\subsection{Thermal conductivity}
We measure and check the thermal conductivity $\kappa$ using the two experimental setups
described in section \ref{Exper}.
\subsubsection{With the dedicated electronics}
\label{dedi}

\begin{figure}[htbp]
\centering
\includegraphics[width=7.0cm]{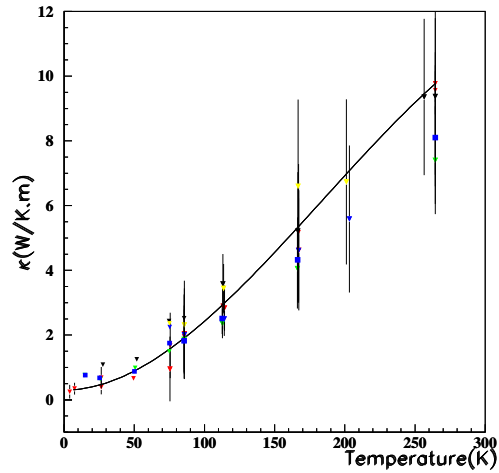}
\caption{\label{Kappa1} {Thermal conductivity $\kappa$
as a function of the temperature for seven Carbon fibers. The full line corresponds to a fit according to
eq. (\ref{kappaeq})}}
\end{figure}

The first step of the analysis using the dedicated electronics is to estimate the resistance
variation of the fibers as a function of the thermalisation temperature for very small voltages
applied on their edges (typically under 100mV), heating the fiber about $10 K$ above $T_0$. 
For such a voltage $V_a$, we measure the amplitude $V_f$ of the signal between $t=0$ and
its asymptotic value, and the difference $\Delta V_{r}$ of the voltage signals
measured at the edges of the same fiber when pulsed in series with the resistance $r=2.65\Omega$
and without $r$ (cf. Figure \ref{Elecfibre2}). The variation of the resistance induced by a tension
$V_a$ applied on the fibers is given by:
\begin{eqnarray}
\Delta R_a = {V_f \over \Delta V_{r}}\ r
\end{eqnarray}
We deduce the value of the mean temperature to which the fiber is heated inverting the polynomial
function used to fit the R(T) behaviour of the fibers presented in section \ref{Rdet}.
For instance, above 40K, the resistance is proportional to the temperature (cf. section \ref{Rdet})
and one gets from (\ref{deltmes}) and (\ref{mualp}):
\begin{eqnarray}
\kappa = - {{V_a}^2 L \alpha \over 12 S \Delta R_a}
\end{eqnarray}
The results are shown on figure \ref{Kappa1} for seven superimposed fibers. $\kappa$ is well described
in this temperature range by the parametrisation:

\begin{eqnarray}
\label{kappaeq}
\kappa &=& \kappa_0 + \kappa_2 T^2 + \kappa_3 T^3
\end{eqnarray}

 the full line curve corresponds to a fit of data with:

 $\kappa_0\simeq 0.3 \pm  0.1 \hbox{W/Km}$,
$\kappa_2\simeq (2.5   \pm  0.8) 10^{-4}
{\hbox{W/K}^3\hbox{m}}$, and $\kappa_3\simeq (-5. \pm  3.)\ 10^{-7} \ {\hbox{W/K}^5\hbox{m}}$.

These results are in good agreement with other measurements on carbon fibers\cite{7}. 

The small slope of $\kappa(T) $ at low temperature makes the fiber signal very stable even if $T_0$ is not stabilized. For instance, with $T_0$ around $4 K$ with variation of the order
of $100 mK$, the fiber signal varies by less than $10^{-3}$.

\subsubsection{Bolometers data}

\begin{figure}[htpb]
\centering
\includegraphics[width=6.0cm]{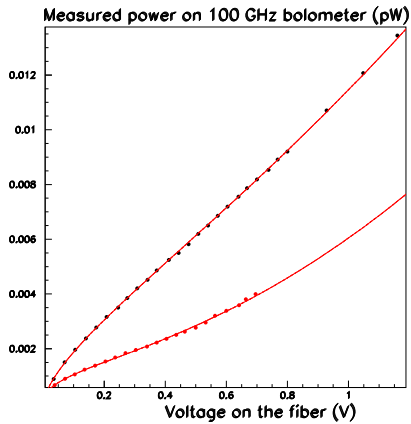}
\caption{Incident power measured by a 100GHz Planck-HFI bolometer
(in pW) - without corrections for optical efficiency of the
cold optics of the instrument -  as a function of the voltage
applied at the edges of the fibers for 2 different fibers:
a 1mm length fiber (in black) and a 0.6mm one (in red). The
superimposed fitted function corresponds to the paramaterization
given in equation (\ref{param}).}
\label{kappa}
\end{figure}

In order to cross-check this result, one makes use of the Planck HFI instrument.
We pulse the fibers with square signals at approximately $1\ Hz$, and compute the amplitude $S_{bolo}$
of the resulting synchroneous signal measured on the bolometers (translated in terms of incident power
on the detectors). $S_{bolo}$ is proportional to the integral of the temperature increase over the length
of the fiber
\footnote{One assumes that, being in the Rayleigh Jeans domain, the emission spectrum of
the fiber is proportional to T.}. Table \ref{spectre} gives the measured values of the flux
(in pW) with the bolometers for the fibers installed on the mirror and pulsed
with 1V amplitude signals\footnote{These values have to be multiplied by a factor
350 if one wants to estimate the flux when the fiber is directly facing
the entrance of the horn of the cold optics in front of the bolometers.}.  

\begin{table}[htbp]
   \begin{center}
\begin{tabular}{|l|l|l|l|l|l|}
   \hline
    & \bf 100GHz & \bf 143 GHz & \bf 217 GHz & \bf 353GHz & \bf 545 GHz \\\hline
Flux($10^{-3}\ \hbox{pW}$)            & 11. & 10. & 21. & 117. & 27. \\
\hline
   \hline
\end{tabular}
\caption{Measured flux for each Planck-HFI frequency band
for the fibers installed on the mirror and pulsed with 1V amplitude
signals. \label{spectre}}
   \end{center}
\end{table}

Figure \ref{kappa} shows the power measured by one of the bolometers
(at 100GHz) for two ``mirror'' fibers (a small 0.6mm fiber and a longer 1mm one) as a function of the
voltage applied on the fiber.

These results can be understood in the light of the framework given in section \ref{frame}.
 In the permanent regime, using (\ref{kappaeq})
to parametrize $\kappa(T)$ and neglecting $R(T)$ dependence, one  gets from equation (\ref{maineq}):
\begin{eqnarray}
{\partial^2  \over  \partial x^2} \left({\kappa_0 T + {\kappa_2 T^3 \over 3} + {\kappa_3 T^4 \over 4}}
\right)
 &=&
-{V^2\over L S R}  \ .
\end{eqnarray}

Since $T=T_0$ for $x=\pm L/2$, one obtains after integration:
\begin{eqnarray}
\label{kappasol}
\kappa_0 T(x)+{\kappa_2\over 3} T^3(x)+{\kappa_3\over 4} T^4(x) &=& A(x)+\gamma
\end{eqnarray}
where:
$$ A(x) ={V^2\over 2 L S R}(L^2/4-x^2) {\rm\ ,and\ }
\gamma = \kappa_0T_0+{\kappa_2\over 3} {T_0}^3+{\kappa_3\over 4} {T_0}^4
\ . $$

Since here $T_0\approx 2K$, it can be neglected with respect to T. Hence,
 $\gamma = 0$.

$A(x)$ simplifies to $A(x) = { V^2 / (8 \mu)} (1-(2x/L)^2)$. Hence, the
solution $T(x)$ is only a function of $V$ and the reduced variable
$\xi = 2x/L$.

For smal $V$ values, the first term dominates the left-hand side of (\ref{kappasol}), that reduces to
(\ref{tx}), and $T(x)$ has a parabolic shape. At medium $V$ values, $T(x)$ solution on most part of
the fiber implies mainly the $T^3$ term, and $T(x)$ has a flatter profile with a $\root 3 \of{1-\xi}$ shape.

At high $V$ value, the negative $\kappa_3$ term plays a higher role and is responsible for the increase in the
$T(x)$ slope. 

The phenomenological expression:
\begin{eqnarray}
\label{param}
S_{bolo} &\propto& G_0 V^2 + {G_2 \over 3} V^{2/3} + {G_3 \over 4} V^{1/2}
\end{eqnarray}

is found to fit accurately both experimental (see figure \ref{kappa}) and simulated data
(see section \ref{simu}).

Here, the applied voltages are higher that the ones used in the previous section,
still the model fits accurately the data and both results are in very good agreement.

\subsection{Time constants and Calorific capacitance}

\begin{figure}[htpb]
\includegraphics[width=7.cm]{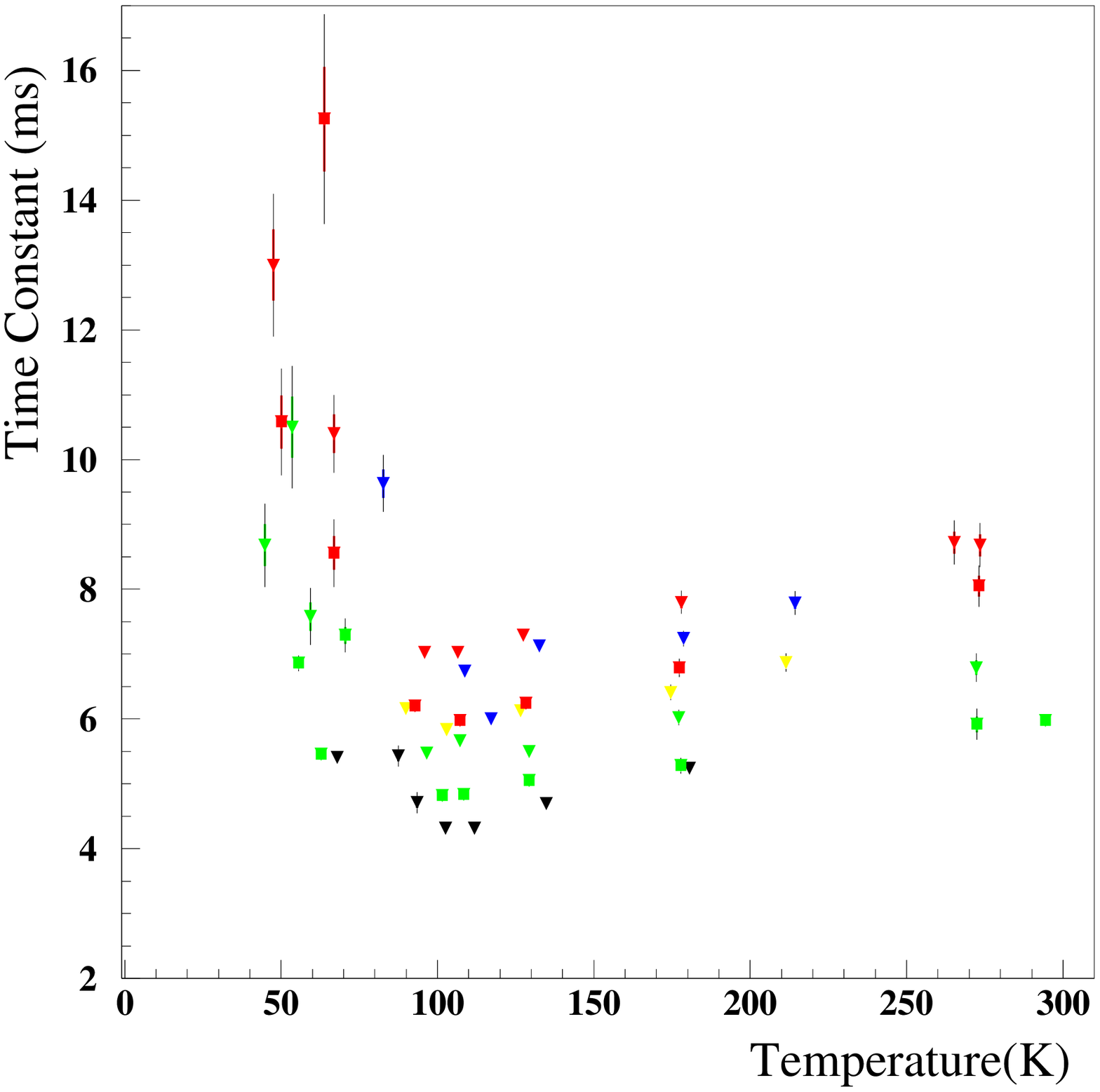}  
\includegraphics[width=7.cm]{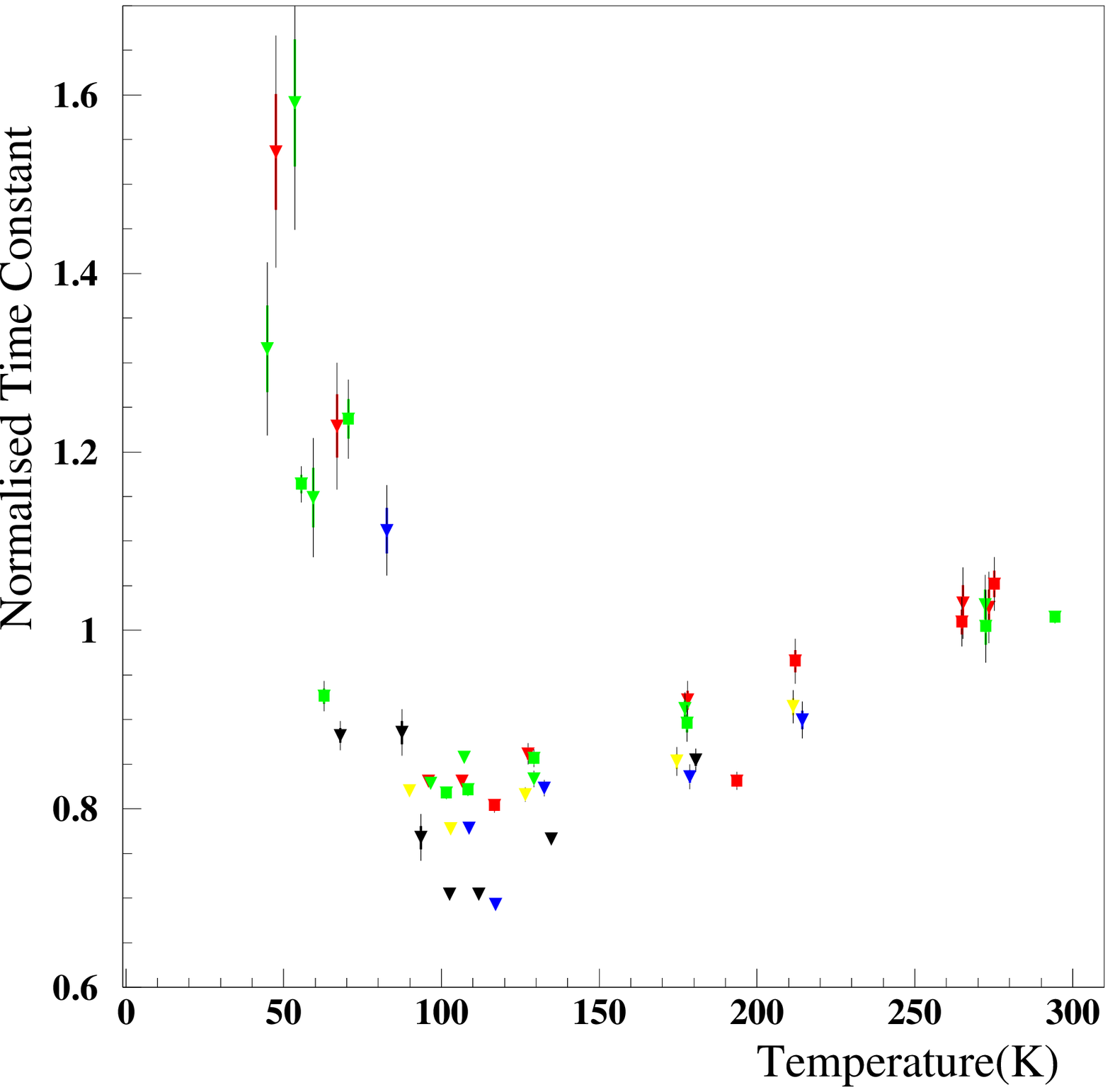}
\caption{On the left: Time constant (in ms) as a function of the thermalisation
temperature. On the right: same data but normalized to the mean value obtained
for each fiber with $T\ge 250K$. }
\label{cterenorm}
\end{figure}

With the dedicated electronics of section (\ref{descr_dedi}), and applying only small
voltages to the fibers, one can extract the fiber time constant by fitting the exponential
behaviour of $V_1(t)$ in a wide range of thermalisation temperature $T_0$. The
results on $\tau$ are illustrated on figure \ref{cterenorm}: on the left hand side the
data are in ms while on the right there are normalized to the mean value obtained
for each fiber with $T\ge 250K$,
showing that this behaviour is coherent from one fiber to the others.

In the temperature range we explore, the fiber time constants present 
a minimum around $100 K$, and rise slowly up to $300 K$. The higher 
time constants measured for a temperature smaller than 50K are 
responsible for long decay times at the end of the pulse.
During the
Planck HFI calibration \cite{Calib}, this drawback has been corrected for using
a small permanent current on the fiber in addition to the pulse. 
This permanent current allows to maintain
most of the fiber at temperatures where the time constant 
remains small, and we recover a fiber decay
time in the same range as the rise time.

\begin{figure}[htbp]
\centering
\includegraphics[width=14.0cm]{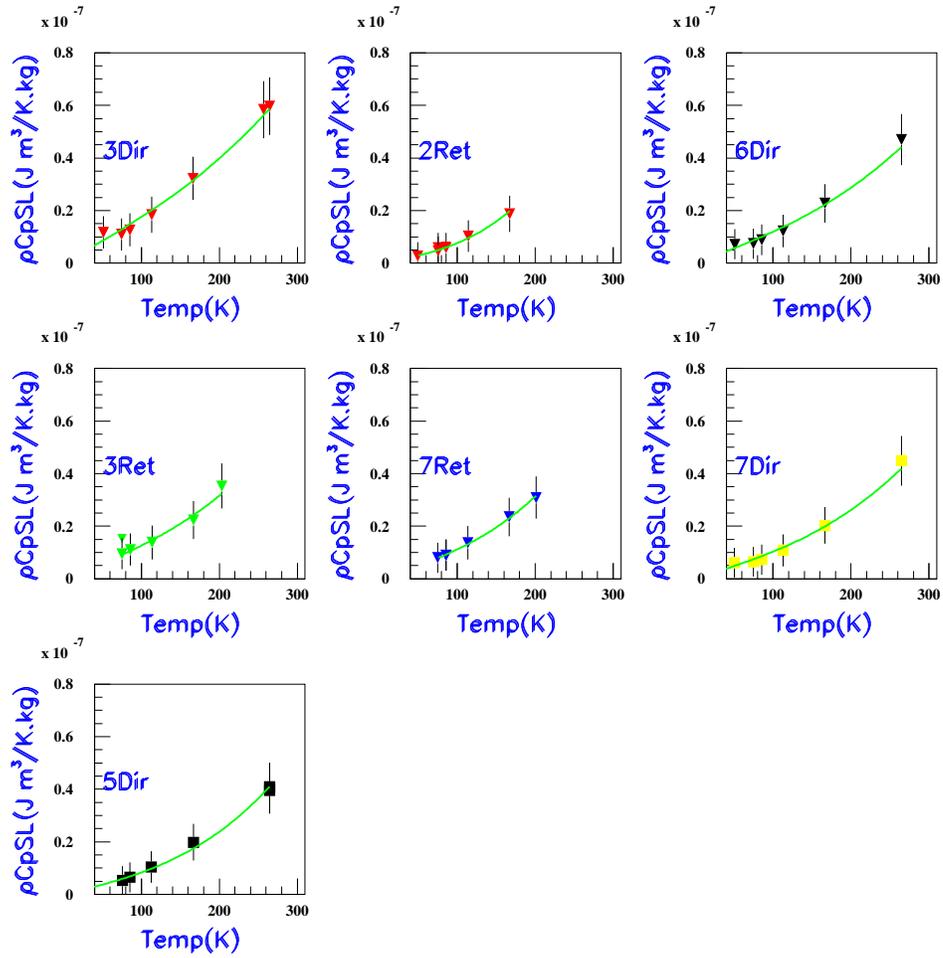}
\caption{\label{Cp} {Calorific capacitance $C_p$ 
as a function of the temperature for several Carbon fibers.}}
\end{figure}

$C_p$ can be deduced from the measurement of the time constant and from the $\kappa$ modelisation
through relation (\ref{tau1}). Figure \ref{Cp} shows individual results for seven fiber calorific
capacitance. Parametric adjustments of the standard type $C_p/T = C_{p0} + T^2 C_{p2}$ are superimposed.
A good qualitative agreement is met with expected values for Carbon fiber at low temperature\cite{5}.

\section{Simulation}
\label{simu}
\begin{figure}[htbp]
\centering
\includegraphics[width=8.0cm]{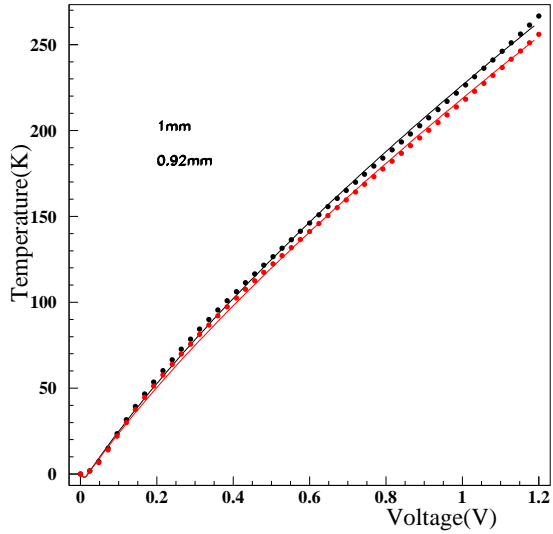}
\caption{\label{Ttens} {Simulation of the mean temperature of the
fiber as a function of the amplitude of the applied voltage for
several fibers lengths.}}
\end{figure}

We developped a simple simulation code solving numerically equation \ref{maineq}. Using the
Crank Nicholson scheme \cite{2} and introducing the fitted dependence on the temperature
of R, $\kappa$  and $C_p$, we can compute the behaviour of the mean temperature
of the fiber as a function of the input voltage: it is shown on  figure \ref{Ttens} which reproduces the results
measured on the bolometers (cf. Figure \ref{kappa}). The parameters extracted from data with some approximations
are therefore coherent with the simulations.

\section{Conclusion}
This article describes the physical parameters of the carbon fiber illuminators used in the calibration of the 
Planck-HFI instrument. We have shown that, thanks to their low time constant (
$\le 10ms$ for 1mm
fibers at 2K), they can be used in a pulsed regime without introducing 
any electrical parasitic signal on HFI's bolometers, and that
their emission spectrum gives a significant amount of signal in the submm domain.

We have detailed an analysis of the temperature dependence of the resistance, the calorific capacitance $C_p$,
and the thermal conductivity $\kappa$, of theses fibers based on measurements from 300 to 1.7K. We end
up showing the good agreement between simulations of the 1D heat equations making use of the extracted 
parameters ($R(T),\ C_p(T)$ and $\kappa(T)$) and data measured of the 100mK HFI's bolometers.

This well understood picture of the fibers behaviour makes them a very usefull tool for
FIR instrument calibration.

{\noindent {\bf Aknowledgments:}
We wish to aknowledge 
J.C. Vanel and C. Rosset for their help in our first attempt to
cool down the fibers,
B. Maffei and R. Sudiwala for providing us with material and
manpower at Cardiff,  
J.P. Torre for his disponibility and his setup,  
O. Perdereau, J. Haissinski and S. Plaszczynski for usefull discussions
and help, and F. Pajot, P. Lami and the Saturne Cryostat team for
the Planck-HFI calibration. \\

This work has been funded by CNES as part of LAL contribution to HFI, under contract Nº 737/CNES/01/8961/00. 

}

\newpage

\end{document}